\documentstyle[12pt,rotate]{article}

\input epsf.sty

\newcommand{\inclfig}[2]{\mbox{\epsfxsize=#1cm \epsfbox{#2.ps}}}
\newcommand{\insertfig}[2]{\mbox{\epsfxsize=#1cm \epsfbox{#2.eps}}}
\newcommand{\g}{{\sl g}}

\newcommand{\J}{{\cal J}}
\newcommand{\Z}{{\cal Z}}

\newcommand{\D}{{\cal D}}
\newcommand{\K}{{\cal K}}

\def\1{\hbox{{1}\kern-.25em\hbox{l}}}

\begin{document}
\begin{titlepage}

\centerline{\large \bf Fine structure of spectrum of}
\centerline{\large \bf twist-three operators in QCD.}

\vspace{10mm}

\centerline{\bf A.V. Belitsky\footnote{Alexander von Humboldt Fellow.}}

\vspace{10mm}

\centerline{\it Institut f\"ur Theoretische Physik, Universit\"at
                Regensburg}
\centerline{\it D-93040 Regensburg, Germany}

\vspace{30mm}

\centerline{\bf Abstract}

\hspace{0.8cm}

We unravel the structure of the spectrum of the anomalous dimensions
of the quark-gluon twist-3 operators which are responsible for the
multiparton correlations in hadrons and enter as a leading contribution
to several physical cross sections. The method of analysis is bases
on the recent finding of a non-trivial integral of motion for the
corresponding Hamiltonian problem in multicolour limit which results
into exact integrability of the three-particle system. Quasiclassical
expansion is used for solving the problem. We address the chiral-odd
sector as a case of study.

\vspace{6cm}


\vspace{0.5cm}


\end{titlepage}

%%%%%%%%%%%%%%%%%%%%%%%%%%%%%%%%%%%%%%%%%%%%%%%%%%%%%%%%%%%%%%%%%%%%%
\section{Introduction.}
%%%%%%%%%%%%%%%%%%%%%%%%%%%%%%%%%%%%%%%%%%%%%%%%%%%%%%%%%%%%%%%%%%%%%

The studies of power suppressed (in a hard momentum scale) phenomena
in physical cross sections give an important insight into underlying
physics of strong interaction since corresponding quantities manifest
multiparton correlations in hadrons and can shed some light on the
complicated dynamics of the hadronic substructure. The most promising
is the study of twist-three contributions in hadronic reactions which
enter as a leading effect in certain asymmetries. The planned and
running experiments will allow to extract these quantities and to face
them with theoretical predictions. However, the question arises about
the scale at which experimental and theoretical predictions are confronted.
This requires the knowledge of the evolution equations which govern
the scale dependence of the corresponding correlation functions ---
a generalization of the familiar one-variable parton densities --- as
well as their solutions, i.e.\ determination of their eigenvalues and
eigenstates. While the former problem is quite straightforward since
it can be easily resolved using standard methods of QCD perturbation
theory, provided a basis of operators mixed under renormalization
is chosen in an appropriate way, and this problem has been tackled by
a number of authors (see e.g.\ reviews \cite{KodTan98,Bel97}). On the
other hand the second problem is by far more complicated since it
requires the solution of the Faddeev-type many-body equations for the
particles with pair-wise interaction. Recently a breakthrough has been
made, however, by the authors of Ref.\ \cite{BraDerMan98} where they
have found an additional non-trivial integral of motion for the problem
at hand. Even so, when the problem acquires an equivalence to an exactly
solvable model the task is proved to be too complicated to be solved
exactly analytically. Therefore, it is welcome to develop an approximate
scheme which could allow for a systematical improvement of approximations
involved. The present paper is devoted to the study along this line for
the chiral-odd twist-three evolution equations, i.e.\ the ones related
to the hadron structure functions $h_L (x)$ and $e (x)$ \cite{JafJi91}.

%%%%%%%%%%%%%%%%%%%%%%%%%%%%%%%%%%%%%%%%%%%%%%%%%%%%%%%%%%%%%%%%%%%%%
\section{Quasi-partonic operators and their evolution.}
%%%%%%%%%%%%%%%%%%%%%%%%%%%%%%%%%%%%%%%%%%%%%%%%%%%%%%%%%%%%%%%%%%%%%

The twist-3 sector is exceptional as compared to even higher twists
since in the former case contrary to the latter we can reduce the
renormalization group analysis to the study of the UV divergencies
of the so-called quasi-partonic operators \cite{BFKL85} which form
a compete basis of functions. The twist of these objects equals
the number of fields which the composite operators are constructed
from. This means that due to the fact that in leading order only
a pair-wise interaction of partons is of relevance, the corresponding
kernels in sub-channels are the familiar twist-2 non-forward
evolution kernels.

The generic form of the evolution equation for twist-3 correlation
functions in the momentum fraction formalism looks like
\begin{equation}
\label{EvolEq}
\mu^2 \frac{d}{d \mu^2} Z (x_1, x_2, x_3)
\!=\! \int \prod_{i=1}^{3} d x'_i \,
\delta \left( \sum_{i = 1}^{3} x'_i - \zeta \right)
\mbox{\boldmath$K$} \left( \{ x_i \} | \{ x'_i \} \right)
Z (x'_1, x'_2, x'_3) ,
\end{equation}
with the constraint $x_1 + x_2 + x_3 = \zeta$ for the momentum fractions
of partons imposed on the both sides of this equation. The variable $\zeta$
stands for the $+$-component of the $t$-channel momentum: the limit
$\zeta = 0$ corresponds to the usual DIS kinematics (and Eq.\
(\ref{EvolEq}) to a generalized DGLAP equation), and $\zeta = 1$ to the
exclusive one (Brodsky-Lepage equation for baryon distribution amplitude).
The kernel $\mbox{\boldmath$K$}$ in leading order has a pair-wise
structure:
\begin{equation}
\mbox{\boldmath$K$} \left( \{x_i\} | \{x'_i\} \right) = \sum_{i < j}
K \left( x_i, x_j | x'_i, x'_j \right)
\delta \left( x_i + x_j - x'_i - x'_j \right) ,
\end{equation}
with $K \left( x_i, x_j | x'_i, x'_j \right)$ being the interaction
kernel of two nearby particles.

For the purposes of the present study we need the quark-anti-quark and
(anti-)quark-gluon kernels with non-contracted Lorentz, Dirac
and colour indices. They can be decomposed into independent structures
as follows
\begin{eqnarray}
\label{qqKernel}
&&\!\!\!\!\!{^{q\bar q}K^{ij;\alpha\beta}_{i'j';\alpha'\beta'}}
\left( x_1, x_2 | x'_1, x'_2 \right) \nonumber\\
&&\!\!\!\!\!\quad
= - \frac{\alpha_s}{2\pi}
\left\{
\frac{1}{4}
\left[
\left( \gamma_- \right)_{\alpha\beta}
\left( \gamma_+ \right)_{\beta'\alpha'}
-
\left( \gamma_- \gamma_5 \right)_{\alpha\beta}
\left( \gamma_+ \gamma_5 \right)_{\beta'\alpha'}
\right]
\left[
\frac{1}{C_A} \delta_{ij} \delta_{j'i'} \,
{^{q \bar q}\! K^V_{(1)}}
+
\frac{1}{T_F} (t^a)_{ij} (t^a)_{j'i'} \,
{^{q \bar q}\! K^V_{(8)}}
\right]
\right. \nonumber\\
&&\!\!\!\!\!\qquad\quad\ \ \
+ \frac{1}{4}
\left.
\left( \sigma^\perp_{-\mu} \right)_{\alpha\beta}
\left( \sigma^\perp_{+\mu} \right)_{\beta'\alpha'}
\left[
\frac{1}{C_A} \delta_{ij} \delta_{j'i'} \,
{^{q \bar q}\! K^T_{(1)}}
+
\frac{1}{T_F} (t^a)_{ij} (t^a)_{j'i'} \,
{^{q \bar q}\! K^T_{(8)}}
\right]
\right\} \left( x_1, x_2 | x'_1, x'_2 \right) ,
\end{eqnarray}
for quark-anti-quark kernel, and
\begin{eqnarray}
\label{qgKernel}
&&{^{qg}K^{ia;\alpha\mu}_{i'a';\alpha'\mu'}}
\left( x_1, x_2 | x'_1, x'_2 \right) \nonumber\\
&&\quad
= - \frac{\alpha_s}{2\pi}
\frac{1}{C_F C_A} (t^a)_{ij} (t^{a'})_{j'i'}
\left\{
- \frac{1}{8}
\left[
\left( \gamma_- \gamma^\perp_\mu \right)_{\alpha\beta}
\left( \gamma_+ \gamma^\perp_{\mu'} \right)_{\beta'\alpha'}
+
\left( \gamma_- \gamma^\perp_\mu \gamma_5 \right)_{\alpha\beta}
\left( \gamma_+ \gamma^\perp_{\mu'} \gamma_5 \right)_{\beta'\alpha'}
\right]
{^{q g}\! K^V_{(3)}}
\right. \nonumber\\
&&\qquad\qquad\qquad\qquad\qquad\qquad\ \
- \frac{1}{4}
\left.
\left( \tau^\perp_{\mu\nu;\rho\sigma}
\gamma_- \gamma^\perp_\nu \right)_{\alpha\beta}
\left( \tau^\perp_{\mu'\nu';\rho\sigma}
\gamma_+ \gamma^\perp_{\nu'} \right)_{\beta'\alpha'}
{^{q g}\! K^T_{(3)}}
\right\} \left( x_1, x_2 | x'_1, x'_2 \right)  \nonumber\\
&&\quad + \dots ,
\end{eqnarray}
for the quark-gluon one. The ellipsis stand for other colour
structures which, however, are irrelevant for our present
consideration. Here as usual the $\pm$-subscripts mean the
projection on two opposite tangents to the light cone, and the
transversity tensor $\tau^\perp_{\mu\nu;\rho\sigma} \equiv
\frac{1}{2} \left( g^\perp_{\mu\rho} g^\perp_{\nu\sigma}
+ g^\perp_{\mu\sigma} g^\perp_{\nu\rho} - g^\perp_{\mu\nu}
g^\perp_{\rho\sigma} \right)$ is constructed from the $2$D metric
tensors of the transverse plane $g^\perp_{\mu\nu} = g_{\mu\nu}
- n_\mu n^\ast_\nu - n_\nu n^\ast_\mu$.

%%%%%%%%%%%%%%%%%%%%%%%%%%%%%%%%%%%%%%%%%%%%%%%%%%%%%%%%%%%%%%%%%%%%%
%            Figure 1
%%%%%%%%%%%%%%%%%%%%%%%%%%%%%%%%%%%%%%%%%%%%%%%%%%%%%%%%%%%%%%%%%%%%%
\begin{figure}[t]
\begin{center}
\vspace{-0.3cm}
\hspace{1cm}
\mbox{
\begin{picture}(0,220)(270,0)
\put(0,-30){\inclfig{18}{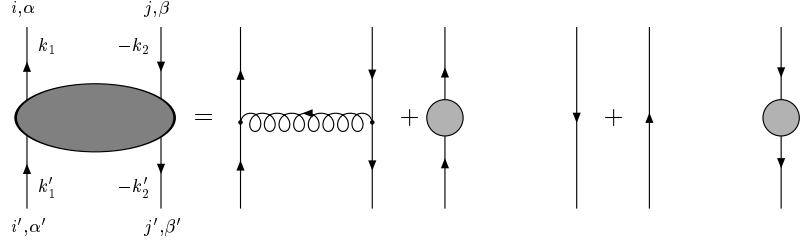}}
\end{picture}
}
\end{center}
\vspace{-5cm}
\caption{\label{qq-kernel} The diagrams contributing to the pair-wise
quark-anti-quark kernel at one-loop order in the light-cone gauge,
$B_+ = 0$. The blobs on the lines stand for the wave function
renormalization counterterm.}
\end{figure}

%%%%%%%%%%%%%%%%%%%%%%%%%%%%%%%%%%%%%%%%%%%%%%%%%%%%%%%%%%%%%%%%%%%%%
%            Figure 2
%%%%%%%%%%%%%%%%%%%%%%%%%%%%%%%%%%%%%%%%%%%%%%%%%%%%%%%%%%%%%%%%%%%%%
\begin{figure}[t]
\begin{center}
\vspace{-0.3cm}
\hspace{1cm}
\mbox{
\begin{picture}(0,220)(270,0)
\put(0,-30){\inclfig{18}{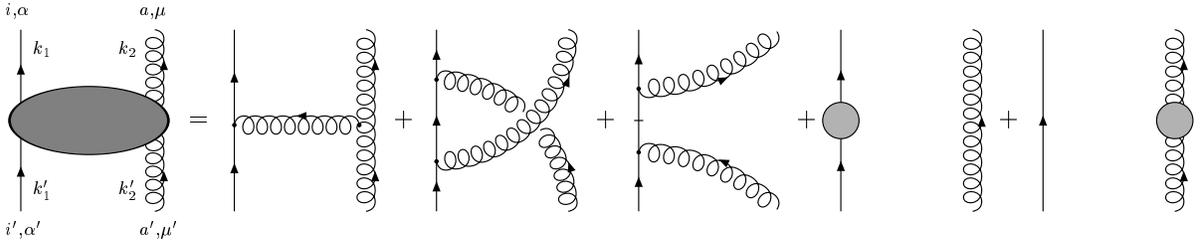}}
\end{picture}
}
\end{center}
\vspace{-5cm}
\caption{\label{qg-kernel} Same as in Fig.\ \protect\ref{qq-kernel}
but for the quark-gluon kernel. The graph with the crossed fermion
propagator corresponds to the contact-type contribution arising from
the use of the Heisenberg equation of motion for the quark field.}
\end{figure}

From diagrams in Fig.\ \ref{qq-kernel} we easily get\footnote{Here
$\Theta^{m}_{i_1 i_2 ... i_n}
(x_1,x_2,...,x_n)=\int_{-\infty}^{\infty}\frac{d\alpha}{2\pi i}
\alpha^m \prod_{k=1}^{n}\left(\alpha x_k -1 +i0 \right)^{-i_k}$.
A detailed discussion of the properties of these functions can be
found in Ref.\ \cite{Bel97}.} (cf.\ \cite{BFKL85,BelMue97a,BelMul98a})
\begin{eqnarray}
&&\frac{1}{C_F}
{^{q \bar q}\! K^V_{(1)}} \left( x_1, x_2 | x'_1, x'_2 \right)
=
\frac{1}{C_F - \frac{C_A}{2}}
{^{q \bar q}\! K^V_{(8)}} \left( x_1, x_2 | x'_1, x'_2 \right) \nonumber\\
&&\qquad
=
\left[
\frac{x_1}{x_1 - x'_1} \Theta^0_{11} (x_1, x_1 - x'_1)
+
\frac{x_2}{x_2 - x'_2} \Theta^0_{11} (x_2, x_2 - x'_2)
+
\Theta^0_{111} (x_1, - x_2, x_1 - x'_1)
\right]_+ , \\
&&\frac{1}{C_F}
{^{q \bar q}\! K^T_{(1)}} \left( x_1, x_2 | x'_1, x'_2 \right)
=
\frac{1}{C_F - \frac{C_A}{2}}
{^{q \bar q}\! K^T_{(8)}} \left( x_1, x_2 | x'_1, x'_2 \right) \nonumber\\
&&\qquad
=
\left[
\frac{x_1}{x_1 - x'_1} \Theta^0_{11} (x_1, x_1 - x'_1)
+
\frac{x_2}{x_2 - x'_2} \Theta^0_{11} (x_2, x_2 - x'_2)
\right]_+
+ \frac{1}{2} \delta (x_1 - x'_1) ,
\end{eqnarray}
and from Fig.\ \ref{qg-kernel} we have (cf.\ \cite{BFKL85})
\begin{eqnarray}
&&{^{q g}\! K^V_{(3)}} \left( x_1, x_2 | x'_1, x'_2 \right) \nonumber\\
&&\qquad
= \frac{C_A}{2}
\left[
\frac{x_1}{x_1 - x'_1} \Theta^0_{11} (x_1, x_1 - x'_1)
\right]_+
+ \frac{C_A}{2}
\frac{x_2}{x'_2}
\left[
\frac{x_2}{x_2 - x'_2} \Theta^0_{11} (x_2, x_2 - x'_2)
\right]_+ \nonumber\\
&&\qquad
+ \frac{C_A}{2} \frac{x_2 + x'_2}{x'_2}
\Theta^0_{111} (x_1, - x_2, x_1 - x'_1)
- \left( C_F - \frac{C_A}{2} \right)
\frac{x_2 - x'_1}{x'_2}
\Theta^0_{111} (x_1, - x_2, x_1 - x'_2) \nonumber\\
&&\qquad
+ C_F \frac{x_2}{x'_2} \frac{x_1}{x_1 + x_2}
\Theta^0_{11} (x_1, - x_2)
+ \frac{1}{8} ( \beta_0 + 5 C_A ) \delta (x_1 - x'_1) , \\
&&{^{q g}\! K^T_{(3)}} \left( x_1, x_2 | x'_1, x'_2 \right) \nonumber\\
&&\qquad
= \frac{C_A}{2}
\left[
\frac{x_1}{x_1 - x'_1} \Theta^0_{11} (x_1, x_1 - x'_1)
\right]_+
+ \frac{C_A}{2}
\frac{x_2}{x'_2}
\left[
\frac{x_2}{x_2 - x'_2} \Theta^0_{11} (x_2, x_2 - x'_2)
\right]_+ \nonumber\\
&&\qquad
- \left( C_F - \frac{C_A}{2} \right)
\frac{x_1}{x'_2} \Theta^0_{11} (x_1, x_1 - x'_2)
+ \frac{1}{8} ( \beta_0 + 5 C_A ) \delta (x_1 - x'_1) .
\end{eqnarray}

With these results at hand the construction of the evolution equation
for the quasi-partonic operators is almost trivial since it requires a
mere evaluation of simple traces of Eqs.\ (\ref{qqKernel},\ref{qgKernel})
with tensor structure of the composite operator under study.

The correlators related to the chiral-odd structure function
$e(x)$ and $h_L(x)$ are given by light-cone Fourier transformation
\cite{BelMue97}
\begin{equation}
\label{CorrelFunct}
Z (x_1, x_2 = - x_1 - x_3, x_3) =
\int \frac{d \kappa_1}{2 \pi} \frac{d \kappa_3}{2 \pi}
e^{i \kappa_1 x_1 + i \kappa_3 x_3}
\langle h |
\frac{1}{2}
\left\{
\Z (\kappa_1, 0, \kappa_3) \pm
\Z ( - \kappa_3, 0, - \kappa_1)
\right\}
| h \rangle
\end{equation}
of the non-local operators
\begin{equation}
\label{oddOperator}
\Z (\kappa_1, \kappa_2, \kappa_3)
= \frac{1}{2} \bar\psi (\kappa_3 n) \sigma^\perp_{\mu +}
\left(
\begin{array}{c}
1 \\
\gamma_5
\end{array}
\right)
t^a \g G^a_{+ \mu} (\kappa_2 n) \psi (\kappa_1 n) .
\end{equation}
The ``$+$" and ``$-$" signs in Eq.\ (\ref{CorrelFunct}) correspond to
$1$ and $\gamma_5$ structures and, i.e.\ to $e$ and $h_L$ functions,
respectively. Thus, we have finally the evolution kernel which governs
the scale dependence of the correlator $Z (x_1, x_2, x_3)$
\begin{eqnarray}
\mbox{\boldmath$K$}^{\rm odd} \left( \{x_i\} | \{x'_i\} \right)
= - \frac{\alpha_s}{2 \pi}
\biggl\{
&&\!\!\!\!\!\!\!\!\!\!
{^{q \bar q}\! K^T_{(8)}} \left( x_1, x_3 | x'_1, x'_3 \right)
\delta (x_1 + x_3 - x'_1 - x'_3) \\
+
&&\!\!\!\!\!\!\!\!\!\!
{^{q g}\! K^V_{(3)}} \left( x_1, x_2 | x'_1, x'_2 \right)
\delta (x_1 + x_2 - x'_1 - x'_2) \nonumber\\
+
&&\!\!\!\!\!\!\!\!\!\!
{^{q g}\! K^V_{(3)}} \left( x_3, x_2 | x'_3, x'_2 \right)
\delta (x_3 + x_2 - x'_3 - x'_2)
- \frac{\beta_0}{4} \delta (x_1 - x'_1) \delta (x_2 - x'_2)
\biggr\} , \nonumber
\end{eqnarray}
where we have added the charge renormalization piece due to presence
of the coupling constant in the definition of the composite operator
(\ref{oddOperator}). Using explicit expressions for pair-wise kernels
we can find that this is exactly the result of Ref.\ \cite{BelMue97}
obtained by a different method.

%%%%%%%%%%%%%%%%%%%%%%%%%%%%%%%%%%%%%%%%%%%%%%%%%%%%%%%%%%%%%%%%%%%%%
\section{Consequences of conformal invariance.}
%%%%%%%%%%%%%%%%%%%%%%%%%%%%%%%%%%%%%%%%%%%%%%%%%%%%%%%%%%%%%%%%%%%%%

It is well-known that the tree-level conformal invariance of the
theory is enough for diagonality of the one-loop anomalous dimensions
matrix of the conformal operators \cite{EfrRad79}-\cite{BelMul99}.
This means that the pair-wise kernels displayed above can be
diagonalized in the basis of the Jacobi polynomials,
$P^{\left( \{ 1, 2 \}, 1 \right)}_j$. Namely, due to the support
properties of the kernels we can write
\begin{eqnarray}
&&\int dx_1 dx_2 \delta (x_1 + x_2 - x'_1 - x'_2)
P^{\left( \{ 1, 2 \}, 1 \right)}_j
\left( \frac{x_1 - x_2}{x_1 + x_2} \right) \,
{^{q \{ \bar q, g \}}\! K^{\mit\Gamma}}
\left( x_1, x_2 | x'_1, x'_2 \right) \\
&&\hspace{10cm} = \frac{1}{2}\,
{^{q \{\bar q, g \}}\! \gamma^{\mit\Gamma}_j}
P^{\left( \{ 1, 2 \}, 1 \right)}_j
\left( \frac{x'_1 - x'_2}{x'_1 + x'_2} \right), \nonumber
\end{eqnarray}
where $j$ is a conserved (in leading order) quantum number related to
the conformal spin of the composite two-particle operators, i.e.\
the eigenvalue of the Casimir operator of the collinear conformal
algebra $su(1,1)$. The eigenvalues of the kernels are\footnote{Note
that from ${\cal N} = 1$ SUSY Ward identities it follows that
${^{q \bar q} \gamma^T}_j = {^{q g} \gamma^T}_j$ for $j=$even,
${^{q \bar q} \gamma^T}_{j + 1} = {^{q g} \gamma^T}_j$ for $j=$odd
at leading order and, therefore, as can be easily seen from the
results displayed below in ${\cal N} = 1$ super Yang-Mills theory
the Hamiltonian of interaction of $n$-particles with the same
helicity is equivalent to the one of the $XXX_{s = - 1}$ spin chain.}
\begin{eqnarray}
{^{q \bar q} \gamma^V_{(8)}}_j
&=& \left( C_F - \frac{C_A}{2} \right)
\left( 2 \psi (j + 1) + 2 \psi (j + 3) - 4 \psi (1) - 3 \right), \\
{^{q \bar q} \gamma^T_{(8)}}_j
&=& \left( C_F - \frac{C_A}{2} \right)
\left( 4 \psi (j + 2) - 4 \psi (1) - 3 \right), \\
{^{q g} \gamma^V_{(3)}}_j
&=&
\frac{C_A}{2}
\left(
2 \psi (j + 1) + 2 \psi (j + 4) - 4 \psi (1) \right) \nonumber\\
&&\hspace{4cm} + \left( C_F - \frac{C_A}{2} \right)
\frac{4 \sigma (j)}{(j + 1 )(j + 2)(j + 3)}
+ \frac{1}{4} \left( \beta_0 - 3 C_A \right) , \\
{^{q g} \gamma^T_{(3)}}_j
&=&
\frac{C_A}{2}
\left(
2 \psi (j + 2) + 2 \psi (j + 3) - 4 \psi (1)
\right)
-
\left( C_F - \frac{C_A}{2} \right)
\frac{2 \sigma (j)}{j + 2}
+ \frac{1}{4} \left( \beta_0 - 3 C_A \right) .
\end{eqnarray}
where $\sigma (j) = (- 1)^j$.
Using this we can deduce the following representation of the kernels
\cite{BelMul98a}
\begin{equation}
{^{\phi_1 \phi_2}\! K} (x_1, x_2 | x'_1, x'_2)
= \frac{1}{2}
\sum_{j = 0}^{\infty}
\frac{w (x_1, \nu_1 | x_2, \nu_2)}{\omega_j (\nu_1, \nu_2)}
P^{\left( \nu_2, \nu_1 \right)}_j
\left( \frac{x_1 - x_2}{x_1 + x_2} \right)
{^{\phi_1 \phi_2}\! \gamma_j}\,
P^{\left( \nu_2, \nu_1 \right)}_j
\left( \frac{x'_1 - x'_2}{x'_1 + x'_2} \right),
\end{equation}
where $w (x_1, \nu_1 | x_2, \nu_2) = x_1^{\nu_1} x_2^{\nu_2}$ and
$\omega_j (\nu_1, \nu_2) = \frac{\Gamma (j + \nu_1 + 1) \Gamma
(j + \nu_2 + 1)}{\left( 2j + \nu_1 + \nu_2 + 1 \right) j! \Gamma
(j + \nu_1 + \nu_2 + 1)}$, with $\nu_\ell = d_\ell + s_\ell - 1$, and
$d_\ell$ and $s_\ell$ being canonical scale dimension and spin of the
constituent $\phi_\ell = \{q, g\}$.

In the basis of local operators the above eigenfunctions correspond
to the conformal operators \cite{Mak80,Ohr82}
\begin{equation}
\label{ConformOper}
{\cal O}_{jl} = \phi_2 (i \partial_+)^l
P^{(\nu_2, \nu_1)}_j
\left( \stackrel{\leftrightarrow}{\partial}_+ / \partial_+ \right)
\phi_1,
\end{equation}
where $\phi_\ell$ is an arbitrary local field. These operators form
an irreducible representation in the space of bilinear composite
operators of the collinear conformal group with generators
$\J^+,\J^-,\J^3$. Here the step-up $\J^+ = i {\cal P}_+$, the step-down
$\J^- = \frac{i}{2} \K_-$ and the grade $\J^3 = \frac{i}{2} (\D
+ {\cal M}_{-+})$ operators, constructed from the momentum, ${\cal P}_+$,
the angular momentum, ${\cal M}_{-+}$, the dilatation, $\D$, and the
special conformal, $\K_-$, generators, form the $su(1,1)$
algebra: $[\J^3 , \J^\pm]_- = \pm \J^\pm$, $[\J^+ , \J^-]_- = -2 \J^3$.
The vacuum state is ${\cal O}_{jj}$, $[\J^-, {\cal O}_{jj}]_- = 0$.
The eigenvalues of the Casimir operator $\mbox{\boldmath$\J$}^2 =
\J^3 ( \J^3 - 1 ) - \J^+ \J^-$ define the conformal spin, $J_{12}$,
of the state, i.e.\ $[\mbox{\boldmath$\J$}^2, {\cal O}_{jl}]_-
= J_{12}(J_{12} + 1) {\cal O}_{jl}$ with $J_{12} = j + \frac{1}{2}(\nu_1
+ \nu_2)$.

Therefore, the evolution equation can be reformulated into the
eigenvalue problem for the three-particle system in a basis of
local operators
\begin{equation}
\label{Eigensystem}
{\cal H}_{\rm QCD}\, {\mit\Psi} = {\cal E}_{\rm QCD}\, {\mit\Psi} ,
\end{equation}
with Hamiltonian
\begin{equation}
\label{FiniteNcHamiltonian}
{\cal H}_{\rm QCD}
= {^{q \bar q}\!{\cal H}^T_{(8)}} (\hat J_{13})
+ {^{q g}\!{\cal H}^V_{(3)}} (\hat J_{12})
+ {^{q g}\!{\cal H}^V_{(3)}} (\hat J_{32})
- \frac{\beta_0}{4} ,
\end{equation}
where we have defined ${^{q \bar q}\!\gamma}_{J - 1} =
2\, {^{q \bar q}\!{\cal H}} (J)$ and ${^{q g}\!\gamma}_{J - 3/2}
= 2\, {^{q g}\!{\cal H}} (J)$. The operators $\hat J$'s are formally
determined as solutions of the equation ${\mbox{\boldmath$\hat J$}}^2
= \hat J (\hat J + 1)$.

In the large $N_c$ limit Eq.\ (\ref{FiniteNcHamiltonian}) simplifies
into
\begin{equation}
\label{LargeNcHamiltonian}
{\cal H}
= \frac{N_c}{2}
\left\{
\psi \left( \hat J_{12} - \frac{1}{2} \right)
+ \psi \left( \hat J_{12} + \frac{5}{2} \right)
+ \psi \left( \hat J_{32} - \frac{1}{2} \right)
+ \psi \left( \hat J_{32} + \frac{5}{2} \right)
- 4 \psi (1) - \frac{3}{2}
\right\} .
\end{equation}
Thus, the Hamiltonians (\ref{FiniteNcHamiltonian},\ref{LargeNcHamiltonian})
explicitly manifest the $SU(1,1)$ invariance of the system.

%%%%%%%%%%%%%%%%%%%%%%%%%%%%%%%%%%%%%%%%%%%%%%%%%%%%%%%%%%%%%%%%%%%%%
\section{$\theta$-space.}
%%%%%%%%%%%%%%%%%%%%%%%%%%%%%%%%%%%%%%%%%%%%%%%%%%%%%%%%%%%%%%%%%%%%%

Let us explore in full the consequences of the covariance of the
problem under the conformal transformations. For this we define a
space $V = \{ \theta^k | k = 0,1,\dots,\infty \}$ spanned by the
elements\footnote{It is worth to note that we work in the space
of local operators rather then with the correlation functions of
composite operator, $\mbox{\boldmath${\cal O}$}$, with elementary
fields, $\phi_\ell$, $\langle \mbox{\boldmath${\cal O}$}
\prod_{\ell} \phi_\ell \rangle$, or in the language of Refs.\
\cite{DerMan97,BraDerMan98} the so-called hat-transformed basis.}
\begin{equation}
\label{ThetaToDeriv}
\theta^k \equiv \frac{\partial^k_+ \phi}{\Gamma (k + \nu + 1)} .
\end{equation}
In the representation $[\J^{\pm,3} ,\chi ( \theta )]_- = \hat J^{\pm,3}
\chi ( \theta )$ the generators are
\begin{equation}
\hat J^+ = (\nu + 1) \theta + \theta^2 \frac{\partial}{\partial\theta},
\quad
\hat J^- = \frac{\partial}{\partial\theta},
\quad
\hat J^3 = \frac{1}{2}(\nu + 1) + \theta \frac{\partial}{\partial\theta} ,
\end{equation}
with commutation relations: $[\hat J^3 , \hat J^\pm]_- = \pm \hat J^\pm$,
$[\hat J^+ , \hat J^-]_- = -2 \hat J^3$. For a multi-variable function
$\chi ( \theta_1, \theta_2,\dots,\theta_n )$ the operators are defined
as $\hat J^{\pm,3} = \sum_{\ell = 1}^{n} \hat J^{\pm,3}_\ell$ and the
quadratic Casimir operator is ${\mbox{\boldmath$\hat J$}}^2 = \hat J^3
( \hat J^3 - 1 ) - \hat J^+ \hat J^-$.

Since the spectrum of eigenvalues of the problem (\ref{Eigensystem})
have to be real it means that the Hamiltonian has to be selfadjoint
w.r.t. an appropriate scalar product. We define it as (see \cite{BarRac77}
for definition of group invariant measures)
\begin{equation}
\label{ScalarProduct}
\langle \chi ( \theta_1, \theta_2,\dots,\theta_n ) |
\chi ( \theta_1, \theta_2,\dots,\theta_n ) \rangle
= \int_{{\mit\Omega}} d {\cal M}\
\chi ( \bar\theta_1, \bar\theta_2,\dots,\bar\theta_n )
\chi ( \theta_1, \theta_2,\dots,\theta_n ) ,
\end{equation}
where
\begin{eqnarray}
d {\cal M} \equiv
\prod_{\ell = 1}^{n}
\frac{d \theta_\ell d \bar\theta_\ell}{2 \pi i}
( 1 - \theta_\ell \bar\theta_\ell )^{\nu_\ell - 1}
\qquad\mbox{and}\qquad
{\mit\Omega} = \bigcup^n_{\ell = 1} \{ |\theta_\ell| \leq 1 \}
\nonumber
\end{eqnarray}
and $\bar\theta = \theta^\ast$. Then the following adjoint properties
are obvious
\begin{eqnarray}
\langle \chi ( \theta_1, \theta_2,\dots,\theta_n ) |
\hat J^{\pm,3} \,
\chi ( \theta_1, \theta_2,\dots,\theta_n ) \rangle
&=&
\langle \hat J^{\mp,3} \,
\chi ( \theta_1, \theta_2,\dots,\theta_n )
| \chi ( \theta_1, \theta_2,\dots,\theta_n ) \rangle , \nonumber\\
\langle \chi ( \theta_1, \theta_2,\dots,\theta_n ) |
{\cal H} \,
\chi ( \theta_1, \theta_2,\dots,\theta_n ) \rangle
&=&
\langle {\cal H} \, \chi ( \theta_1, \theta_2,\dots,\theta_n )
| \chi ( \theta_1, \theta_2,\dots,\theta_n ) \rangle . \nonumber
\end{eqnarray}

Now we are ready to address the question of construction of
irreducible representations in $\theta$-space.

%%%%%%%%%%%%%%%%%%%%%%%%%%%%%%%%%%%%%%%%%%%%%%%%%%%%%%%%%%%%%%%%%%%%%
\section{Two-point basis.}
%%%%%%%%%%%%%%%%%%%%%%%%%%%%%%%%%%%%%%%%%%%%%%%%%%%%%%%%%%%%%%%%%%%%%

The highest weight vector in the $\theta$-space depending on two
variables, $\hat J^-_{12}\, \chi (\theta_1 , \theta_2) = 0$, is
realized by translation invariant polynomials\footnote{We use
everywhere the shorthand notation $\theta_{\ell\ell'} \equiv
\theta_\ell - \theta_{\ell'}$.} $\chi (\theta_1 , \theta_2 )
= \theta_{12}^j$. The descendants are constructed by acting with the
step-up operator on the vacuum state: $\left( {\hat J}^+_{12} \right)^k
\theta_{12}^j$. When transformed to the basis of the local operators
(\ref{ThetaToDeriv}) they coincide with Eq.\ (\ref{ConformOper})
up to an overall normalization
\begin{eqnarray}
{\cal O}_{j, j + k} = i^{j + k}
\frac{\Gamma (j + \nu_1 + 1) \Gamma (j + \nu_2 + 1)}{\Gamma (j + 1)}
\left( {\hat J}^+_{12} \right)^k \theta_{12}^j . \nonumber
\end{eqnarray}
The states with unit norm w.r.t.\ the scalar product
(\ref{ScalarProduct}) are
\begin{equation}
{\cal P}_j (\theta_1 , \theta_2)
= n^{-1} (j | \nu_1, \nu_2) \theta_{12}^j ,
\quad\mbox{with}\quad
n^2 (j | \nu_1, \nu_2) = \frac{\Gamma (\nu_1) \Gamma (\nu_2)
\Gamma (j + 1) \Gamma (2 j + \nu_1 + \nu_2 + 1)}{
\Gamma (j + \nu_1 + 1) \Gamma (j + \nu_2 + 1)
\Gamma (j + \nu_1 + \nu_2 + 1)} ,
\end{equation}
so that $\langle {\cal P}_{j'} ( \theta_1, \theta_2 )|
{\cal P}_j ( \theta_1, \theta_2 ) \rangle = \delta_{j'j}$.
The two-particle Casimir operator
\begin{equation}
{\mbox{\boldmath$\hat J$}}_{12}^2
= - \theta_{12}^2
\frac{\partial}{\partial\theta_1}
\frac{\partial}{\partial\theta_2}
- \theta_{12}
\left(
(\nu_1 + 1) \frac{\partial}{\partial\theta_2}
-
(\nu_2 + 1) \frac{\partial}{\partial\theta_1}
\right)
+ \frac{\nu_1 + \nu_2}{2} \left( \frac{\nu_1 + \nu_2}{2} + 1 \right)
\end{equation}
is obviously diagonal in the basis
\begin{equation}
\langle {\cal P}_{j'} ( \theta_1, \theta_2 )|
{\mbox{\boldmath$\hat J$}}^2_{12} | {\cal P}_j ( \theta_1, \theta_2 )
\rangle
= \delta_{j'j}\,
[{\mbox{\boldmath$\hat J$}}^2_{12}]_{jj},\quad\mbox{with}\quad
[{\mbox{\boldmath$\hat J$}}^2_{12}]_{jj} =
\left( j + \frac{\nu_1 + \nu_2}{2}\right)
\left( j + \frac{\nu_1 + \nu_2}{2} + 1 \right) .
\end{equation}

%%%%%%%%%%%%%%%%%%%%%%%%%%%%%%%%%%%%%%%%%%%%%%%%%%%%%%%%%%%%%%%%%%%%%
\section{Three-point basis.}
%%%%%%%%%%%%%%%%%%%%%%%%%%%%%%%%%%%%%%%%%%%%%%%%%%%%%%%%%%%%%%%%%%%%%

In order to construct an orthonormal basis of three-particles
operators we expand them w.r.t.\ the eigenfunctions of the two-point
Casimir operator, say ${\mbox{\boldmath$\hat J$}}^2_{12}$, as follows
\begin{equation}
\label{ThreePointBasis}
{\cal P}_{J; j} ( \theta_1, \theta_2 | \theta_3 )
= \sum_{k = 0}^{J - j}
P_k\, \theta_3^{J - j - k}
\left( {\hat J}^+_{12} \right)^k {\cal P}_j (\theta_1, \theta_2 ).
\end{equation}
From the condition\footnote{In what follows operators without
subscripts stand for three-particle ones while two-particle
operators in subchannels are labeled by corresponding indices.}
$\hat J^- {\cal P}_{J; j} ( \theta_1, \theta_2 | \theta_3 ) = 0$ we
deduce the expansion coefficients
\begin{equation}
P_k = N^{-1} (J, j | \nu_1, \nu_2, \nu_3)
\frac{(-1)^k \Gamma (J - j + 1)}{\Gamma (k + 1) \Gamma (J - j - k + 1)}
\frac{\Gamma (2 j + \nu_1 + \nu_2 + 2)}{
\Gamma (2 j + k + \nu_1 + \nu_2 + 2)} ,
\end{equation}
where the factor
\begin{equation}
N^2 (J, j | \nu_1, \nu_2, \nu_3)
= \frac{\Gamma (\nu_3) \Gamma (J - j + 1)
\Gamma (2 j + \nu_1 + \nu_2 + 2 )
\Gamma (2 J + \nu_1 + \nu_2 + \nu_3 + 2)}{
\Gamma (J - j + \nu_3 + 1) \Gamma (J + j + \nu_1 + \nu_2 + 2)
\Gamma (J + j + \nu_1 + \nu_2 + \nu_3 + 2) }.
\end{equation}
ensures the normalization of the state to unity
$\langle {\cal P}_{J';j'} ( \theta_1, \theta_2 | \theta_3)|
{\cal P}_{J;j} ( \theta_1, \theta_2 | \theta_3) \rangle =
\delta_{j'j} \delta_{J'J}$. From the construction it is obvious
that ${\mbox{\boldmath$\hat J$}}^2_{12}$
and total Casimir operator
\begin{equation}
{\mbox{\boldmath$\hat J$}}^2 =
{\mbox{\boldmath$\hat J$}}^2_{12}
+ {\mbox{\boldmath$\hat J$}}^2_{23}
+ {\mbox{\boldmath$\hat J$}}^2_{13}
+ \sum_{\ell = 1}^{3} \frac{1 - \nu_\ell^2}{4} ,
\end{equation}
which is the sum of the two-particle Casimir operators in
subchannels minus the single-particle ones, are diagonal in the
basis ${\cal P}_{J; j} ( \theta_1, \theta_2 | \theta_3 )$
\begin{eqnarray}
&&\langle {\cal P}_{J'; j'}
( \theta_1, \theta_2 | \theta_3 )
| {\mbox{\boldmath$\hat J$}}^2_{12} |
{\cal P}_{J; j} ( \theta_1, \theta_2 | \theta_3 ) \rangle
= \delta_{J'J} \delta_{j'j}
[{\mbox{\boldmath$\hat J$}}^2_{12} ]_{jj} , \\
&&\langle {\cal P}_{J'; j'}
( \theta_1, \theta_2 | \theta_3 )
| {\mbox{\boldmath$\hat J$}}^2 |
{\cal P}_{J; j} ( \theta_1, \theta_2 | \theta_3 ) \rangle
= \delta_{J'J} \delta_{j'j}\, [{\mbox{\boldmath$\hat J$}}^2]_{jj} ,
\end{eqnarray}
where $[{\mbox{\boldmath$\hat J$}}^2]_{jj} = [\hat J_3]_{jj} \left(
[\hat J_3]_{jj} - 1 \right) $ and $[\hat J_3]_{jj} =  J
+ \frac{1}{2}(\nu_1 + \nu_2 + \nu_3 + 3 )$. Since the main quantum
number $J$ (total conformal spin) is conserved we do not display the
dependence on it in matrix elements.

The matrix elements of the remaining generators can be easily evaluated
and the result is
\begin{equation}
\langle {\cal P}_{J'; j'}
( \theta_1, \theta_2 | \theta_3 ) |
{\mbox{\boldmath$\hat J$}}^2_{23}
| {\cal P}_{J; j} ( \theta_1, \theta_2 | \theta_3 ) \rangle
= \delta_{J'J}
\left(
\delta_{j'j} [ {\mbox{\boldmath$\hat J$}}^2_{23} ]_{jj}
+
\delta_{j',j + 1} [ {\mbox{\boldmath$\hat J$}}^2_{23} ]_{j + 1,j}
+
\delta_{j',j - 1} [ {\mbox{\boldmath$\hat J$}}^2_{23} ]_{j - 1,j}
\right)
\end{equation}
where the diagonal part is
\begin{equation}
[ {\mbox{\boldmath$\hat J$}}^2_{23} ]_{jj}
= \frac{1}{2}
\left(
[ {\mbox{\boldmath$\hat J$}}^2 ]_{jj}
-
[ {\mbox{\boldmath$\hat J$}}^2_{12} ]_{jj}
-
\sum_{\ell = 1}^{3} \frac{1 - \nu_\ell^2}{4}
+
\frac{\nu_2^2 - \nu_1^2}{4}
\frac{1}{ [ {\mbox{\boldmath$\hat J$}}^2_{12} ]_{jj} }
\left(
[ {\mbox{\boldmath$\hat J$}}^2 ]_{jj}
+
\frac{1 - \nu_3^2}{4}
\right)
\right) ,
\end{equation}
and the non-diagonal elements are
\begin{equation}
[ {\mbox{\boldmath$\hat J$}}^2_{23} ]_{j + 1,j}
=
\frac{{\cal N} (J, j)}{{\cal N} (J, j + 1)}
(j + 1)(J - j + \nu_3) ,
\qquad
[ {\mbox{\boldmath$\hat J$}}^2_{23} ]_{j - 1,j}
=
\frac{{\cal N} (J, j - 1)}{{\cal N} (J, j)}
j(J - j + \nu_3 + 1) ,
\end{equation}
with ${\cal N} (J, j) \equiv N (J, j | \nu_1, \nu_2, \nu_3)
n (j | \nu_1, \nu_2)$. So that
$[ {\mbox{\boldmath$\hat J$}}^2_{23} ]_{j + 1,j} =
[ {\mbox{\boldmath$\hat J$}}^2_{23} ]_{j,j + 1}$.

%%%%%%%%%%%%%%%%%%%%%%%%%%%%%%%%%%%%%%%%%%%%%%%%%%%%%%%%%%%%%%%%%%%%%
\section{Integral of motion and its quantization.}
%%%%%%%%%%%%%%%%%%%%%%%%%%%%%%%%%%%%%%%%%%%%%%%%%%%%%%%%%%%%%%%%%%%%%

On top of the conformal invariance there exists another hidden
symmetry of the system (\ref{LargeNcHamiltonian}). The beautiful
finding made by the authors of Ref.\ \cite{BraDerMan98} is the
identification of an additional conserved charge of the three-particle
problem\footnote{See Ref.\ \cite{BraDerKorMan99} for an exhaustive
treatment of the three-quark problem in the context of solution of
the Brodsky-Lepage evolution equation for baryon distribution
amplitudes.} described by the Hamiltonian (\ref{LargeNcHamiltonian})
in the limit of large number of colours,
\begin{equation}
\label{Charge}
{\cal Q}_T
= [ {\mbox{\boldmath$\hat J$}}^2_{12},
{\mbox{\boldmath$\hat J$}}^2_{23} ]_+
- \frac{9}{2}
\left\{
{\mbox{\boldmath$\hat J$}}^2_{12}
+ {\mbox{\boldmath$\hat J$}}^2_{23}
\right\} ,
\end{equation}
which is the hermitian operator and commutes with the Hamiltonian
(\ref{LargeNcHamiltonian}) $[{\cal H}, {\cal Q}_T]_- = 0$.
Everywhere we have to put $\nu_1 = \nu_3 = 1$, $\nu_2 = 2$.

The existence of this additional charge leads to complete integrability
of the system (the number of integrals of motion equals the number of
degrees of freedom). This allows to reduce the complicated eigenfunction
problem for the Hamiltonian (\ref{LargeNcHamiltonian}) to the
more simple one for the ${\cal Q}_T$:
\begin{equation}
\label{QTeigenEq}
{\cal Q}_T {\mit\Psi} = q_T {\mit\Psi} .
\end{equation}
We will look for the solution in the form of expansion w.r.t.\
the three-point basis (\ref{ThreePointBasis}), i.e.\
\begin{equation}
{\mit\Psi} = \sum_{j = 0}^{J} {\mit\Psi}_j {\cal P}_{J;j} .
\end{equation}
Its main advantage comes from the fact that the matrix elements of
the charge possess only three non-zero diagonals:
\begin{equation}
\langle {\cal P}_{J'; j'}
( \theta_1, \theta_2 | \theta_3 ) |
{\cal Q}_T
| {\cal P}_{J; j} ( \theta_1, \theta_2 | \theta_3 ) \rangle
= \delta_{J'J}
\left(
\delta_{j'j} [ {\cal Q}_T ]_{jj}
+
\delta_{j',j + 1} [ {\cal Q}_T ]_{j + 1,j}
+
\delta_{j',j - 1} [ {\cal Q}_T ]_{j - 1,j}
\right) ,
\end{equation}
with
\begin{eqnarray}
&&{[ {\cal Q}_T ]}_{jj}
=
2 \, [ {\mbox{\boldmath$\hat J$}}^2_{12} ]_{jj}
[ {\mbox{\boldmath$\hat J$}}^2_{23} ]_{jj}
- \frac{9}{2} \,
\left\{
[ {\mbox{\boldmath$\hat J$}}^2_{12} ]_{jj}
+ [ {\mbox{\boldmath$\hat J$}}^2_{23} ]_{jj}
\right\}, \\
&&{[ {\cal Q}_T ]}_{j + 1,j}
=
\left\{
[ {\mbox{\boldmath$\hat J$}}^2_{12} ]_{j + 1,j + 1}
+ [ {\mbox{\boldmath$\hat J$}}^2_{12} ]_{jj}
- \frac{9}{2}
\right\} \,
[ {\mbox{\boldmath$\hat J$}}^2_{23} ]_{j + 1,j} ,
\end{eqnarray}
and ${[ {\cal Q}_T ]}_{j + 1,j} = {[ {\cal Q}_T ]}_{j,j + 1}$.
Therefore, the above equation (\ref{QTeigenEq}) leads to a three-term
recursion relation for ${\mit\Psi}_j$ which when solved gives the
expansion coefficients as well as quantized values of ${\cal Q}_T$.
The former is of the form
\begin{equation}
\label{RecRel}
\left( {[ {\cal Q}_T ]}_{j,j} - q_T \right) {\mit\Psi}_j
+ {[ {\cal Q}_T ]}_{j,j - 1} {\mit\Psi}_{j - 1}
+ {[ {\cal Q}_T ]}_{j,j + 1} {\mit\Psi}_{j + 1} = 0 .
\end{equation}
with the boundary conditions ${\mit\Psi}_{- 1} = {\mit\Psi}_{J + 1} = 0$.
The solution for this recursion relation which satisfies the
boundary conditions exists provided the following constraint is
fulfilled
\begin{equation}
\label{ExistOfSol}
\left( {[ {\cal Q}_T ]}_{j,j} - q_T \right)^2
\leq
4 {[ {\cal Q}_T ]}_{j,j + 1} {[ {\cal Q}_T ]}_{j,j - 1} .
\end{equation}

Using the results found so far we can easily deduce from the
Eq.\ (\ref{ExistOfSol}) the critical eigenvalues of the spectrum
of the operator ${\cal Q}_T$ for large conformal spin $J$ of the
three particle system. Substituting the asymptotics for the matrix
element of the operator ${\cal Q}_T$ in the three-point basis
$\left. {[ {\cal Q}_T ]}_{j,j} \right|_{J\to\infty}
= 2 \left. {[ {\cal Q}_T ]}_{j,j + 1} \right|_{J\to\infty} = J^4 \tau^2
(1 - \tau)^2$ where $\tau \equiv j/J$, in Eq.\ (\ref{ExistOfSol})
we get $0 \leq q_T \leq 2 J^4 \tau^2 (1 - \tau^2)$. The maximum
is achieved for $\tau_{\rm max} = 1/\sqrt{2}$. Thus the recurrence
relation possesses the rising solution for $0 \leq j \ll
\frac{1}{\sqrt{2}} J$ and decreasing one for $\frac{1}{\sqrt{2}} J \ll
j \leq J$. We get finally an estimate for the quantized values of the
${\cal Q}_T$ for $J \to \infty$. Namely, the eigenvalues lie in the strip
\begin{equation}
0 \leq q_T / J^4 \leq \frac{1}{2}.
\end{equation}
Note that numerically the upper bound on the spectrum is attained at
very high $J$'s. For instance, for $J = 10^3$ we have $1.00199 \cdot
10^{-6} \leq q_T/J^4 \leq 0.50531$.

%%%%%%%%%%%%%%%%%%%%%%%%%%%%%%%%%%%%%%%%%%%%%%%%%%%%%%%%%%%%%%%%%%%%%
%                      Figure 3
%%%%%%%%%%%%%%%%%%%%%%%%%%%%%%%%%%%%%%%%%%%%%%%%%%%%%%%%%%%%%%%%%%%%%
\begin{figure}[t]
\unitlength1mm
\begin{center}
\vspace{1.5cm}
\hspace{-2cm}
\begin{picture}(100,155)(0,0)
\put(0,60){\insertfig{12}{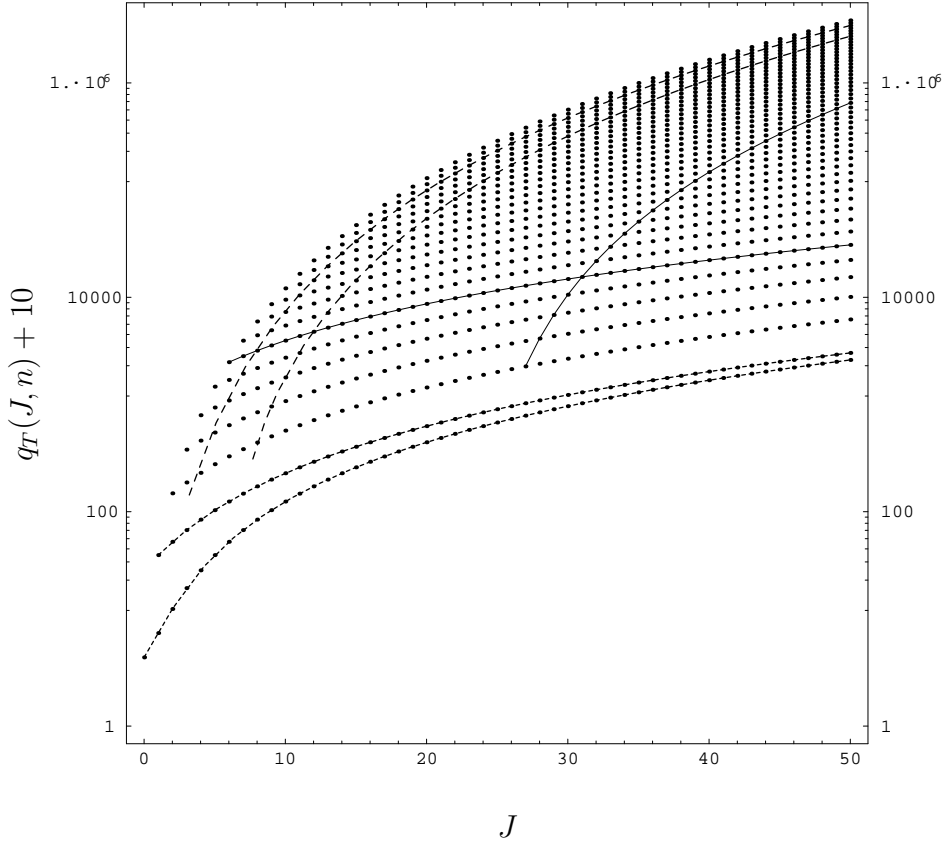}}
\put(60,60){$J$}
\put(-5,110){\rotate{$q_T (J, n) + 10$}}
\end{picture}
\end{center}
\vspace{-6.5cm}
\caption{\label{QTcharge} The spectrum of the conserved charge $q_T$.
Sample trajectories from two sets: for $n = 25$ (counts from above,
$n = 0,1,\dots$) and $m = 7$ (counts from below, $m = 1,2,\dots$) are
shown by solid curves. Long-dashed lines correspond to the analytical
formulae (\protect\ref{WKBexp},\protect\ref{1stWKB}) with $n = 2,6$.
The two exact solutions (\protect\ref{ExactQT}) separated from the rest
of the spectrum by a finite gap are demonstrated by short-dashed lines.}
\end{figure}

We are now in a position to estimate non-leading in $1/J^\ell$
corrections which will generate the fine structure of the spectrum.
Let us consider the upper bound of the spectrum. Since the maximal
eigenvalues are achieved for $j_{\rm max} = \frac{1}{\sqrt{2}} J$
we consider a small vicinity of $j_{\rm max}$:
$j = \frac{1}{\sqrt{2}}\left( J + \lambda \sqrt{J}\right)$. We
look for the expansion of the eigenfunctions ${\mit\Psi}_j
\equiv {\mit\Phi} (\lambda)$ and the charge $q_T$ in series
\begin{equation}
\label{WKBexp}
{\mit\Phi} (\lambda) = \sum_{\ell = 0}^{\infty}
{\mit\Phi}_{(\ell)} (\lambda) J^{- \ell/2} ,
\qquad\mbox{and}\qquad
q_T (J, n) = J^4 \sum_{\ell = 0}^{\infty} q_T^{(\ell)}(n) J^{- \ell},
\quad\mbox{with}\quad q_T^{(0)} = \frac{1}{2} ,
\end{equation}
respectively. Then the difference equation (\ref{RecRel}) is
replaced by the sequence of coupled differential equations
\begin{eqnarray}
\label{Hermite}
&&{\mit\Phi}_{(0)}^{\prime\prime} (\lambda)
+ 4 \left( 6 - q_T^{(1)} - 2 \lambda^2 \right)
{\mit\Phi}_{(0)} (\lambda) = 0, \\
\label{HermiteComb}
&&{\mit\Phi}_{(1)}^{\prime\prime} (\lambda)
+ 4 \left( 6 - q_T^{(1)} - 2 \lambda^2 \right)
{\mit\Phi}_{(1)} (\lambda)
= 8 \lambda \left( 4 \sqrt{2} - 6 + \lambda^2 \right)
{\mit\Phi}_{(0)} (\lambda), \\
&&\dots . \nonumber
\end{eqnarray}
The solution of the first one, which satisfies the boundary conditions,
i.e.\ ${\mit\Phi} (\pm\infty) = 0$, is expressed by Hermite polynomials
\begin{equation}
{\mit\Phi}_{(0)} (\lambda)
= H_n \left( \sqrt{2 \sqrt{2}} \lambda \right) e^{- \sqrt{2} \lambda^2} ,
\end{equation}
giving the quantized values of $q_T^{(1)}$
\begin{equation}
\label{1stWKB}
q_T^{(1)} (n) = \sqrt{2} \left( 3 \sqrt{2} - \frac{1}{2} - n \right),
\quad n =0,1,\dots .
\end{equation}
Here for a given $J$, $0 \leq n \leq J$ but the Eq.\ (\ref{Hermite})
has been derived in the approximation $n \ll J$ and thus we can describe
the upper part of the spectrum only. The comparison of this
approximation with the eigenvalues evaluated numerically is shown
in Fig.\ \ref{QTcharge}. The agreement is reasonable already with the
first non-leading correction taken into account. We can go to the
region $n \sim J$ provided a large enough number of terms is kept in
Eqs.\ (\ref{WKBexp}).

While for low part of the spectrum one can find two exact solutions:
\begin{equation}
\label{ExactQT}
q_T^{\rm exact-1} (J) = - \frac{53}{8} + (J + 1)^2,
\qquad
q_T^{\rm exact-2} (J) = - \frac{53}{8} + (J + 5)^2,
\end{equation}
separated from the rest of the spectrum by a finite gap of order
$\Delta q_T (J) \propto 0.7 J^2 + 40 J$ for large $J$'s.

%%%%%%%%%%%%%%%%%%%%%%%%%%%%%%%%%%%%%%%%%%%%%%%%%%%%%%%%%%%%%%%%%%%%%
\section{Eigen-energy of three-particle system.}
%%%%%%%%%%%%%%%%%%%%%%%%%%%%%%%%%%%%%%%%%%%%%%%%%%%%%%%%%%%%%%%%%%%%%

The eigenfunctions of the integral of motion ${\cal Q}_T$
simultaneously diagonalize the Hamiltonian (\ref{LargeNcHamiltonian})
and, therefore, the energy of the system is\footnote{We have used
the permutation symmetry of the quark fields so that $P_{13} {\mit\Psi}
= e^{i \varphi} {\mit\Psi}$, and $\sum_j {\mit\Psi}_j {\cal P}_{J;j}
(\theta_3, \theta_2| \theta_1) = e^{i \varphi} \sum_j {\mit\Psi}_j
{\cal P}_{J;j} (\theta_1, \theta_2| \theta_3)$ since $P_{13}
{\cal P}_{J;j} (\theta_1, \theta_2| \theta_3) = e^{i \varphi}
{\cal P}_{J;j} (\theta_3, \theta_2| \theta_1)$ where $\varphi = 0,\pi$.}
\begin{equation}
\label{EnergyAnalytic}
{\cal E} (J, n) = \frac{N_c}{2}
\left\{ 2
\left( \sum_{j = 0}^{J} |{\mit\Psi}_j|^2 \right)^{-1}
\sum_{j = 0}^{J} \epsilon(j)
|{\mit\Psi}_j|^2 - \frac{3}{2}
\right\} ,
\end{equation}
with $\epsilon (j) = \psi (j + 1) + \psi (j + 4) + 2 \gamma_E$.
The upper limit on the spectrum can easily be read from the fact
that the function ${\mit\Psi}_j$ is peaked at $j_{\rm max}$ and thus
${\cal E}^{\rm max} (J) = N_c/2 \left\{ 2 \epsilon \left( J/\sqrt{2}
\right) - \frac{3}{2} \right\}$. But this estimate contains
non-leading terms as well.

%%%%%%%%%%%%%%%%%%%%%%%%%%%%%%%%%%%%%%%%%%%%%%%%%%%%%%%%%%%%%%%%%%%%%
%                      Figure 4
%%%%%%%%%%%%%%%%%%%%%%%%%%%%%%%%%%%%%%%%%%%%%%%%%%%%%%%%%%%%%%%%%%%%%
\begin{figure}[t]
\unitlength1mm
\begin{center}
\vspace{0.5cm}
\hspace{-1cm}
\begin{picture}(100,155)(0,0)
\put(0,60){\insertfig{10.4}{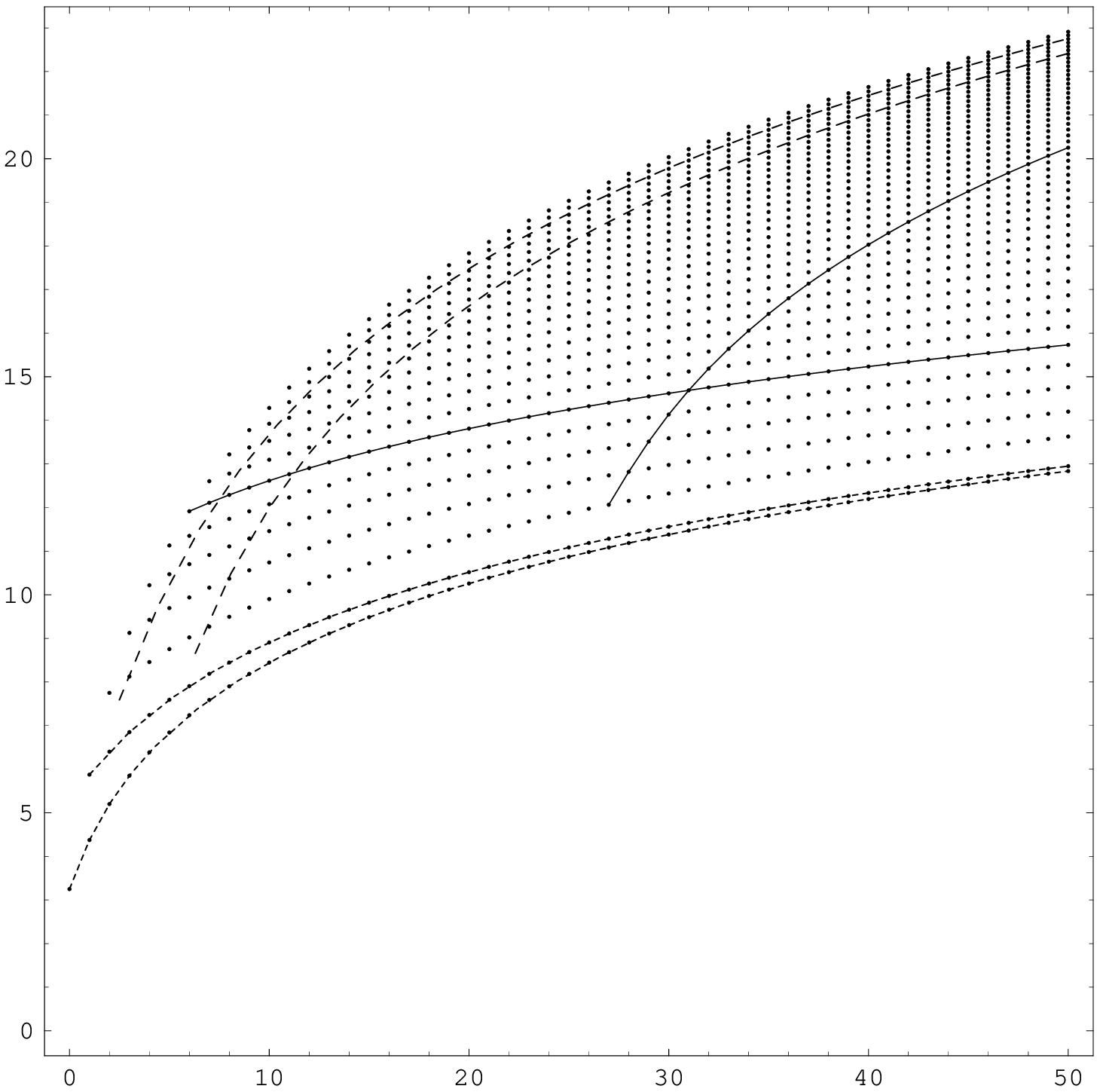}}
\put(53,55){$J$}
\put(-10,110){\rotate{${\cal E} (J, n)$}}
\end{picture}
\end{center}
\vspace{-6cm}
\caption{\label{Energy} Same as in Fig.\ \protect\ref{QTcharge}
but for the spectrum of energy eigenstates. The two lowest exact
trajectories are calculated from Eqs.\ (\protect\ref{ExactE}).}
\end{figure}

The energy can be evaluated consistently in WKB approximation as
\begin{equation}
\label{EnergyWKB}
{\cal E} (J, n) = \frac{N_c}{2}
\left\{
{\cal E}^{(0)} (J)
+ \sum_{\ell = 1}^{\infty} {\cal E}^{(\ell)}(n) J^{- \ell}
\right\} .
\end{equation}
Combining Eqs.\ (\ref{EnergyWKB}) and (\ref{EnergyAnalytic})
with known eigenfunctions allow to determine the coefficients
${\cal E}^{(\ell)}$. However, to evaluate next-to-leading WKB
correction ${\cal E}^{(1)}$ we have to find ${\mit\Phi}_{(1)}$
from Eq.\ (\ref{HermiteComb}) as well. The solution of this
equation is given by
\begin{equation}
\label{NLOeigenfunction}
{\mit\Phi}_{(1)} (\lambda) = \int_{- \infty}^{\infty}
d \lambda' \, G (\lambda, \lambda')
{\cal J} (\lambda') .
\end{equation}
Here the source ${\cal J} (\lambda) = 8 \lambda \left( 4 \sqrt{2} - 6
+ \lambda^2 \right) {\mit\Phi}_{(0)} (\lambda)$ and the Green
function of the homogeneous equation is
\begin{equation}
G (\lambda, \lambda') = \frac{1}{2\sqrt{2\sqrt{2}}}
\sum_{m = 0}^{\infty}
\frac{e^{- \sqrt{2} (\lambda^2 + \lambda'^2)}}{\sqrt{\pi} 2^m m! (n - m)}
H_m \left( \sqrt{2 \sqrt{2}}\lambda \right)
H_m \left( \sqrt{2 \sqrt{2}}\lambda' \right).
\end{equation}
We have omitted, however, in Eq.\ (\ref{NLOeigenfunction}) the solution
of the homogeneous equation which might be added due to presence of
a zero-mode (for $m=n$), since in the energy the corresponding contribution
disappears being an odd function of $\lambda$. Substituting finally the
expansion (\ref{WKBexp}) in Eq.\ (\ref{EnergyAnalytic}) approximating
the sum by the integral in the vicinity of the maximum of ${\mit\Psi}_j$
we find the energy to ${\cal O}(J^{-1})$ accuracy ${\cal E} = N_c \bigl\{
\epsilon^{(0)} - 3/4 + J^{-1} ( 2 \int \epsilon^{(1)} {\mit\Phi}_{(0)}
{\mit\Phi}_{(1)} + \int \epsilon^{(2)} {{\mit\Phi}_{(0)}}^2 ) /
( \int {{\mit\Phi}_{(0)}}^2 ) \bigr\}$ with $\epsilon^{(0)}
= 2 \ln J - \ln 2 + 2 \gamma_E$, $\epsilon^{(1)} = 2 \lambda$,
$\epsilon^{(2)} = 4 \sqrt{2} - \lambda^2$. This gives us finally
\begin{equation}
{\cal E}^{(0)} (J)
= 4 \ln J + 4 \gamma_E - 2 \ln 2 - \frac{3}{2} ,
\qquad
{\cal E}^{(1)} (n) = 2 \sqrt{2}
\left( 3 \sqrt{2} - \frac{1}{2} - n \right) .
\end{equation}
The restriction to this approximation for the energy works somewhat
worse then for the $q_T$-charge (see Fig.\ \ref{Energy}) but can be
improved routinely. From these results we derive the following
expression for the energy via the integral of motion at large
conformal spin
\begin{equation}
{\cal E} = \ln \left( q_T/2 \right) + {\rm const} + {\cal O} (J^{-2}),
\end{equation}
which should be compared with ${\cal E}_{XXX} = 2 \ln q + {\rm const}
+ {\cal O}( J^{-2} )$ for the closed $XXX$ quantum spin chain \cite{Kor96}.

The lowest trajectories corresponding to the solutions (\ref{ExactQT})
reads
\begin{eqnarray}
\label{ExactE}
{\cal E}^{\rm exact-1} (J) = \frac{N_c}{2}
\left\{
2 \psi (J + 3) + 2 \gamma_E - \frac{1}{2} - \frac{1}{J + 3}
\right\}, \\
{\cal E}^{\rm exact-2} (J) = \frac{N_c}{2}
\left\{
2 \psi (J + 3) + 2 \gamma_E - \frac{1}{2} + \frac{3}{J + 3}
\right\} . \nonumber
\end{eqnarray}
And these are exactly the anomalous dimensions found in Ref.\
\cite{BBKT97} (see also \cite{KoiNis97,BelMue97}).

%%%%%%%%%%%%%%%%%%%%%%%%%%%%%%%%%%%%%%%%%%%%%%%%%%%%%%%%%%%%%%%%%%%%%
\section{Outlook and conclusions.}
%%%%%%%%%%%%%%%%%%%%%%%%%%%%%%%%%%%%%%%%%%%%%%%%%%%%%%%%%%%%%%%%%%%%%

%%%%%%%%%%%%%%%%%%%%%%%%%%%%%%%%%%%%%%%%%%%%%%%%%%%%%%%%%%%%%%%%%%%%%
%                      Figure 5
%%%%%%%%%%%%%%%%%%%%%%%%%%%%%%%%%%%%%%%%%%%%%%%%%%%%%%%%%%%%%%%%%%%%%
\begin{figure}[t]
\unitlength1mm
\begin{center}
\vspace{0.5cm}
\hspace{-1.5cm}
\begin{picture}(100,155)(0,0)
\put(0,60){\insertfig{10.4}{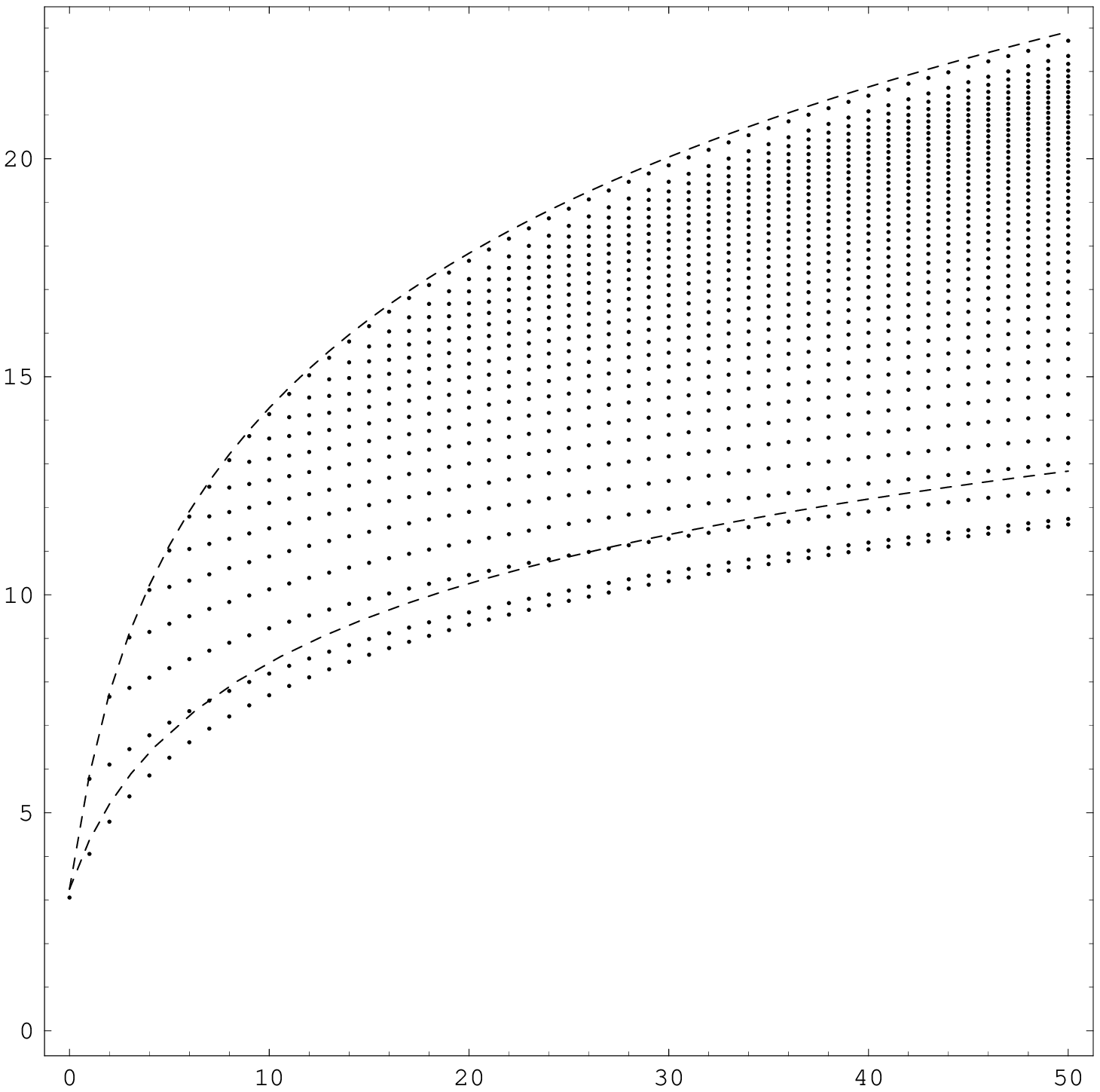}}
\put(53,55){$J$}
\put(-10,110){\rotate{${\cal E}_{\rm QCD} (J, n)$}}
\end{picture}
\end{center}
\vspace{-6cm}
\caption{\label{FiniteNc} The spectrum of eigenvalues of Hamiltonian
(\protect\ref{FiniteNcHamiltonian}) compared to the range of
eigenvalues in the multicolour limit (dashed curves).}
\end{figure}

To conclude we have constructed a convenient orthonormal three-particle
basis in the space of local composite operators. Making use of it we have
found that the allowed eigenvalues of the ``hidden" integral of motion
${\cal Q}_T$ and the energy ${\cal E}$ of the system lie in the strips
for large $J$'s: $J^2 \leq q_T \leq \frac{1}{2} J^4$ and $2 \ln J \leq
{\cal E} \leq 4 \ln J$ with the two lowest levels given by Eqs.\
(\ref{ExactQT},\ref{ExactE}) separated by finite gaps from the rest of
the spectrum. The first non-leading WKB corrections which describe the
fine structure of the spectra have been found. An improvement of these
approximations could systematically be achieved by taking into account
further ${\cal O} \left( J^{-\ell} \right)$ $(\ell \geq 2)$ terms in
the recursion relation (\ref{RecRel}). However, this procedure becomes
non-trivial for higher WKB corrections. Another fruitful way for
determination the energy of the system would be the identification of
the corresponding three-site integrable open spin chain model \cite{Skl88}
with the charge (\ref{Charge}) defined from the auxiliary transfer matrix
$\hat t (\lambda) = {\rm Tr}\, \left( T (\lambda) K^- (\lambda)
T^{-1} ( - \lambda) K^+ (\lambda) \right)$ with boundary operators
$K^\pm$ satisfying the Sklyanin equation and auxiliary monodromy matrix
$T (\lambda) = \prod_{\ell = 1}^{3} L_\ell (\lambda)$, given by the product
of Lax operators $L (\lambda) = \lambda \1 + \sigma_+ \hat J^- -
\sigma_- \hat J^+ + \sigma_3 \hat J^3$, which fulfills the
ordinary Yang-Baxter equation. However, since the Hamiltonian
${\cal H}$ (\ref{LargeNcHamiltonian}) of the system is essentially
quantum and does not enter into the expansion of $\hat t (\lambda)$ in
rapidity $\lambda$, one has to construct the fundamental transfer
matrix as well. Once the functional Bethe ansatz is developed it would
allow to construct the WKB solution \cite{Kor96,Kor97} of the
corresponding Baxter equation and as a result to deduce higher WKB
corrections to the $q_T$ in a more economic way together with the
corresponding energy. Unfortunately the theory of open quantum spin
chains \cite{OpenChain} is far from being developed to the level of
periodic models.

Of course, for real QCD case (finite $N_c$) the integrability of
the system is violated by $1/N_c^2$ corrections in Eq.\
(\ref{FiniteNcHamiltonian}). These effects manifest themselves
in the phenomenon of generation of a mass gap between the $n = 0$
trajectory and the rest of the spectrum (see Fig.\ \ref{FiniteNc}).

\vspace{1cm}

We would like to thank V. Braun for the talk on the evolution of
baryon distribution amplitudes \cite{BraDerKorMan99} at the
Workshop "Structure functions and hadronic wave functions" held
in Bad Honnef (Germany) in the middle of December 1998 which
inspired the author to start the present study. We are grateful
to G. Korchemsky and A. Manashov for very helpful correspondence
and the Alexander von Humboldt Foundation for support.

\end{document}